# Laser induced suppression of transmission in magnetically strained black phosphorus

by

R. Biswas[1], R. Dey[2] and C. Sinha[1, 3]

[1]Department of Physics, P. K. College, Contai, Purba Medinipur, West Bengal- 721401, India

[2]Department of Physics, Govt. G. D. College, Susunia, Purulia, West Bengal-723131, India.

[3]Department of Theoretical Physics, Indian Association for the Cultivation of Science, Jadavpur-700032, India

**Abstract:**

Charge transport through a rectangular vector potential barrier modulated by a continuum laser in monolayer phosphorene is studied theoretically in the ballistic regime along the line of Floquet formalism. Laser free transmission profile displays strong directional behavior exhibiting collimation depending on the incident energy and width of the barrier. However, the application of laser, polarized along the zig-zag direction, creates a sharp anti-resonance in the transmission spectrum and reveals a strong light matter interaction due to broken symmetry in presence of the magnetic vector potential. Transmission properties through a vector barrier are found to be sensitive particularly for lower frequency and higher intensity of the laser. For a thin barrier, the laser assisted conductance is suppressed remarkably in contrast to its oscillatory nature for a thicker one.

KeyWords: Phosphorene; Laser; Vector barrier; Conductance

## Introduction:

After the major breakthrough of the experimental realization of graphene in 2004 [1, 2], some other 2D materials that have created sensation in the world of nano technology are silicine, $MOS_2$, germanene etc. and then one of the most technologically potential candidate black phosphorus. Of late, the layered black phosphorus (BP) has created tremendous attention to the researchers particularly owing to its unique anisotropic electronic properties [3-5] and layer dependent direct band gaps [6], in contrast to the widely studied graphene. In fact, the zero band gap of graphene, due to its low energy linear dispersion imposes severe limitations in the field of digital electronics. The low ON/OFF ratio of graphene based switches leads to a decline of graphene application. In order to circumvent this problem Scientists were in search of some other 2D materials having finite band gap resulting in the discovery of the aforesaid materials (Silicene, Germanene, Phosphorene, etc.) which possess intrinsic thickness dependent band gap (in contrast to graphene) originating due to relatively large spin orbit interaction. Such band gap can be tuned by external electric field. The band gap of BP increases when the thickness of the material decreases from bulk to few layers and eventually monolayer [7-8].

Of late, layered BP has gained tremendous potentiality in multidisciplinary fields of nano and opto-electronics due to its unique electronic and opto-electronic properties. In the bulk form BP is a weak Vander Waals – bonded layered material where each layer form a puckard surface due to $sp^3$ hybridization of the 3s and 3p atomic orbitals unlike the graphene where layers are perfectly flat and hence possesses isotropic band honeycomb structure [9]. In particular, the field effect transistor (FET) based on a few layer BP is found to have an ON/OFF ratio of $10^5$ [10] and carrier mobility at room temperature as high as $10^3$ cm$^2$/V.s that make BP a favourable material for the next generation electronics [11-12].

One of the most recent 2D materials, the phosphorene is a monolayer of black phosphorus and is the most stable allotrope of phosphorus. The most distinguishing feature of phosphorene that differentiates it with the widely used popular material graphene is its anisotropic puckered honey comb band structure. Due to this puckerring, the unit cell of phosphorene contains four atoms [13] unlike the two atoms in graphene. The puckered arrangement of phosphorene along the arm chaired direction, while a bilayer structure along the zig-zag direction has drawn vast attention of researchers compared to other 2D layered materials. Further, phosphorene, due to its natural band gap, might be useful in switching between insulating and conducting states by tuning the band gap. The high aniosotropic property of phosphorene also enables it to exhibit interesting direction dependent optical and transport properties.

Phosphorene also possesses a quasi-one dimensional (1D) excitonic nature [14] in sharp contrast with the properties of other 2D materials, e.g., graphene and transition metal dichalcogenide semiconductors (TMDS) [15]. In particular, monolayer phosphorene might be of great interest in investigating fundamental phenomenon such as 2D quantum confinement and also in exploring many body interactions [15].

Further, phosphorene is found to posses several properties (e.g., fundamental band gap, anisotropic effective transport mass along the armchair to zig zag directions, high carrier mobility, etc. ) ideally suited for the high performance tunnel field effect transistor (TFET) applications and is superior to the TMDS in this respect. A recent work reported the monolayer BP TFETs excellent device performance along the armchair direction [16].

Application of external fields on 2D materials is very important particularly because they can control and modify both the transport and the optical properties of these materials significantly. Such studies can also provide theoretical insight regarding the interaction of charge carriers with the external fields in low dimensions. In particular, in case of phosphorene, use of static electric field has been quite successful in controlling the carrier mobility and the anisotropic band transport [17-18]. As regards the effect of external magnetic field on phosphorene, the anisotropic Landau levels were found to generate at low field regime whereas a self-similar Hofstadter butterfly spectra were noted in the high field regime [19]. The orientation of the magnetic barrier was found to have strong influence on the conductance of the system and a suppression of the latter was noted for a barrier parallel to the armchair direction [20]. Further, in a separate work, the strength, orientation and the periodic length of the magnetic barrier superlattice were found to have remarkable impact on the anisotropic energy dispersion in phosphorene [21].

Finally, it is not surprising that the material phosphorus should demonstrate unique optical properties as the word 'Phosphorus' itself originates from the Greek name that carries the meaning of "light bringer". Regarding the light-matter interaction, phosphorene acts as a marvelous test bed for such studies by virtue of its tunable band gap varying from 0.3 eV in bulk to 2.1 eV in the monolayer. Realization of high performance photo-electronic devices could be possible in the nanoscale using phosphorene because of strong light matter interaction [22]. From the semiconducting phase a monolayer phosphorene may switch over to either metallic or topological insulating phases via Lifshitz transition by the application of a laser field [23].

In order to have an insight about how the above interaction is modified by the external magnetic field, it is inevitable to study either the transport or optical properties in presence of both the magnetic field as well as laser. This leads to the study of magneto-optical transport properties in phosphorene mono and multi-layers [24]. As is well known, theoretical studies of such laser driven tunneling transport through the phosphorene are highly demanding particularly for the fabrication of nano opto-electronic devices. On account of the layer dependent electronic band gap of phosphorene, the 2D material exhibits strong light-matter interaction in the visible and infrared frequencies and hence acts as a marvelous test bed for such studies. For better optimization of the phosphorene based opto electronic devices, it is highly desirable to have deep theoretical insight into light matter interaction in phosphorene in different perspectives. Some theoretical works regarding this are already available in the literature [21 - 24]. In particular, for the advancement of nano-optoelectronic devices, the study of laser driven quantum tunneling phenomena including various states on magnetic fields are highly demanding. The present study motivates us such an unexplored direction particularly in respect of magneto laser tunneling in a monolayer phosphorene. Some previous theoretical studies have been made on magneto optical studies of mono and multilayer phosphorene.

The present work addresses the quantum magneto-optic tunneling transport through a monolayer phosphorene vector barrier irradiated by laser. Although it is reported that due to symmetry, the application of radiation polarized along the y-direction has no effect on photonic absorption [25] in phosphorene, here we have shown that the presence of time reversal symmetry breaking magnetic vector potential term presence in the Hamiltonian may lead to photonic transition and thereby modifies the tunneling conductivity in phosphorene vector barrier.

**Theory:**

The present work deals with the electron tunneling through a pair of delta function magnetic barriers [26, 27] in monolayer phosphorene. The corresponding vector potential (Fig. 1) for the magnetic field is given by $\vec{A}_m(x) = (0, A_m, 0)$ in region II (of length '$D$') and zero elsewhere (in regions I and III), $A_m$ being the magnitude of the constant magnetic vector potential in units of $B_0L_0$ where the magnetic length is given by $L_0 = \sqrt{\frac{\hbar}{eB_0}}$. Since a piecewise constant (of length '$D$') vector potential produces a pair of two oppositely directed delta function magnetic barrier (with inter barrier separation '$D$'), the structure under consideration is termed as a rectangular vector barrier (RVB) of width '$D$'.

In general the energy dispersion of phosphorene should be described by a four band model in the tight binding framework. However it can also be expressed by a two band model where the unit cell contains two phosphorus atoms, one in the upper layer while the other in the lower. The effective 2x2 Hamiltonian in region – II based on the tight binding model is given by [6, 28]

$$H_0 = \begin{pmatrix} E_c + \alpha k_x^2 + \beta(k_y + eA_m)^2 & \gamma k_x \\ \gamma k_x & E_v - \lambda k_x^2 - \eta(k_y + eA_m)^2 \end{pmatrix} \qquad (1)$$

where $E_c$ and $E_v$ are the conduction and valence band energies and $k_x$ and $k_y$ are the x and y components of the momentum in the plane of the phosphorene. $(\alpha, \beta)$ $and$ $(\lambda, \eta)$ are constants depending on the effective mass of the electron and hole respectively. $\gamma$ is the coupling constant between the two atoms of the reduced unit cell (known as the effective interband coupling) consisting of two atoms in the upper and lower layer of phosphorene. The corresponding parameters are taken from Ref. [29]. The Hamiltonian equation (1) is based on the effective two band tight binding model [6] where the reduced unit cell of phosphorene contains only two atoms ($A_1$, $B_1$) or ($A_2$, $B_2$) either in the upper or in the lower layer (by virtue of the $D_{2h}$ point invariance [13]).

Let us now apply an external monochromatic, linearly polarized, continuum laser field in region II, perpendicular to the plane of the phosphorene (x-y plane) polarized along the y-direction, given by the time dependent vector potential in the dipole approximation [30, 31] $\vec{A}_l(t) = (0, A_l cos\omega t, 0)$, $A_l$ and $\omega$ being the amplitude and frequency of laser potential. As an approximation, we consider the effect of laser within the barriers (region – II) only, which could be quite justified for heavily doped emitter and collector leads [32]. The corresponding time dependent Hamiltonian for region II can be written by

$$H(t) = \begin{pmatrix} E_c + \alpha k_x^2 + \beta(k_y + eA_m + eA_l cos\omega t)^2 & \gamma k_x \\ \gamma k_x & E_v - \lambda k_x^2 - \eta(k_y + eA_m + eA_l cos\omega t)^2 \end{pmatrix} \qquad (2)$$

The transport direction (x-axis) is chosen to be perpendicular to both the magnetic field (z-axis) and the laser field (y-axis).

The present calculation is performed in the framework of Floquet theory [31]. By virtue of the periodicity of $H(t)$ in time, the solution of the time dependent Schrödinger equation $H(t)\Psi(\vec{r},t) = i\hbar \frac{\partial \Psi(\vec{r},t)}{\partial t}$ may be written as $\Psi(\vec{r},t) = \varphi(\vec{r},t) \exp(-iEt/\hbar)$, (3) where $\vec{r} = x\hat{\imath} + y\hat{\jmath}$ and $E$ is the Floquet quasi-energy. The wave function $\varphi(\vec{r},t)$ satisfies the equation $\varphi(\vec{r}, t + T) = \varphi(\vec{r},t)$, $T$ being the time period of the external laser field. In the present two band model we consider $\varphi(\vec{r},t) = \big(\varphi_a(\vec{r},t), \varphi_b(\vec{r},t)\big)^T$, where the superscript $T$ corresponds to the transpose of the matrix. In view of eqns. (1) - (3) we obtain a set of two coupled differential equations (in dimensionless form) as follows

$$\left[E_c - \alpha \frac{\partial^2}{\partial x^2} + \beta(k_y + A_m)^2 + \beta(k_y + A_m)A_l cos\omega t\right] \varphi_a(x,t) - i\gamma \frac{\partial}{\partial x}\varphi_b(x,t) = \left(E + i\frac{\partial}{\partial t}\right)\varphi_a(x,t) \quad 4(a)$$

$$\left[E_v + \lambda \frac{\partial^2}{\partial x^2} - \eta(k_y + A_m)^2 - \eta(k_y + A_m)A_l cos\omega t\right] \varphi_b(x,t) - i\gamma \frac{\partial}{\partial x}\varphi_a(x,t) = \left(E + i\frac{\partial}{\partial t}\right)\varphi_b(x,t) \quad 4(b)$$

We consider all the potentials to be uniform along the y- direction (the y-component of the wave function $\sim e^{ik_y y}$) and the terms containing higher order of $A_l$ are neglected which is quite legitimate in the low intensity regime of the laser field. For the sake of mathematical simplicity, we consider similar time dependence for the wave function of the two atoms (A&B) in the reduced unit cell of phosphorene, i.e.,

$\varphi_a(x,t) = \varphi_a(x)f(t)$ and $\varphi_b(x,t) = \varphi_b(x)f(t)$ (5)

In view of eqns.4(a), 4(b) and (5) one can find that the spatial part of the wave functions satisfy the relations

$$\left[E_c + \beta(k_y + A_m)^2 - \alpha \frac{\partial^2}{\partial x^2}\right] \varphi_a(x) - i\gamma \frac{\partial}{\partial x}\varphi_b(x) = E_x \varphi_a(x) \qquad 6(a)$$

$$\left[E_v - \eta(k_y + A_m)^2 + \lambda \frac{\partial^2}{\partial x^2}\right] \varphi_b(x) - i\gamma \frac{\partial}{\partial x}\varphi_a(x) = E_y \varphi_b(x) \qquad 6(b)$$

Regarding the time part, $f(t)$ in eqn. (5) satisfies the following two differential equations of the same form;

$$\left[E + i\frac{\partial}{\partial t} + \beta(k_y + A_m)A_l \cos \omega t\right] f(t) = E_x f(t) \qquad 7(a)$$

$$\left[E + i\frac{\partial}{\partial t} - \eta(k_y + A_m)A_l \cos \omega t\right] f(t) = E_y f(t) \qquad 7(b)$$

Here $E_x$ and $E_y$ are arbitrary constants independent of space and time.

Eqns. 7(a) and 7(b) reveal that $f(t)$ possesses two different solutions of the form

$$f(t) \sim Exp\left[-i\left\{n\omega t + \delta(k_y + A_m)\frac{A_l}{\omega}\sin\omega t\right\}\right] \qquad 7(c)$$

but with different band parameters, e.g., $\delta = \beta$ (for conduction band) and $\delta = -\eta$ (for valence band). The time periodicity of the function $f(t)$ leads to $E_x = E_y = E + n\omega$, $n$ being the number of photons exchanged, e.g., positive (negative) for absorption (emission) processes and zero for no photon process. The justification for the band dependent function $f(t)$ could be well understood if we first neglect the inter-band coupling term ($\gamma$) and then try to find the solution for $f(t)$.

The above identification for $E_x$ and $E_y$ finally leads to a set of two coupled differential equations corresponding to the $m$-th Floquet side band given by,

$$\left[E_c + \beta(k_y + A_m)^2 - \alpha\frac{\partial^2}{\partial x^2}\right]\varphi_a^m(x) - i\gamma\frac{\partial}{\partial x}\varphi_b^m(x) = (E + m\omega)\varphi_a^m(x) \qquad 8(a)$$

$$\left[E_v - \eta(k_y + A_m)^2 + \lambda\frac{\partial^2}{\partial x^2}\right]\varphi_b^m(x) - i\gamma\frac{\partial}{\partial x}\varphi_a^m(x) = (E + m\omega)\varphi_b^m(x) \qquad 8(b)$$

which can be solved exactly for $\varphi_a^m(x)$ and $\varphi_b^m(x)$, where '$m$' now stands for the side band index according to the Floquet theory.

Finally, the form of the wave functions $\varphi_{a,b}^m(x,y,t)$ in the laser driven region corresponding to m-th sideband can be given by,

$$\varphi_a^m(x,y,t) = N\,\varphi_a^m(x)\,e^{ik_y y}\,Exp\left[-i\left\{n\omega t + \beta(k_y + A_m)\frac{A_l}{\omega}Sin\omega t\right\}\right] \qquad 9(a)$$

and

$$\varphi_b^m(x,y,t) = N\,\varphi_b^m(x)\,e^{ik_y y}\,Exp\left[-i\left\{n\omega t - \eta(k_y + A_m)\frac{A_l}{\omega}Sin\omega t\right\}\right] \qquad 9(b)$$

where the $x$- components $\varphi_{a,b}^m(x)$ are given by,

$$\begin{pmatrix}\varphi_a^m(x)\\\varphi_b^m(x)\end{pmatrix} = A_m\begin{pmatrix}1\\\xi_1^m\end{pmatrix}e^{ik_1^m x} + B_m\begin{pmatrix}1\\\xi_2^m\end{pmatrix}e^{ik_2^m x} + C_m\begin{pmatrix}1\\\xi_3^m\end{pmatrix}e^{ik_3^m x} + D_m\begin{pmatrix}1\\\xi_4^m\end{pmatrix}e^{ik_4^m x} \qquad (10)$$

Here $A_m, B_m, C_m$ and $D_m$ are the constant coefficients; $k_i^m$'s are the solutions for the qarktic equation

$$\alpha\lambda(k_i^m)^4 + (k_i^m)^2(\gamma^2 - \alpha X_{mv} + \lambda Y_{mc}) - X_{mv}Y_{mc} = 0$$

where $\quad X_{mv} = \{E_v - \eta(k_y^B)^2 - E_m\}, \; Y_{mc} = \{E_c + \beta(k_y^B)^2 - E_m\} \quad$ and

$\xi_i^m = \frac{\gamma k_i^m}{\left[E_m - E_v + (k_i^m)^2 + \eta(k_y^B)^2\right]}$ with $E_m = E + m\omega$ and $k_y^B = k_y + A_m$.

On the other hand, the solutions for the field free regions (Region – I and Region – III) are quite simple and may be obtained from eqn.(8) with $A_m = 0$. Once the solutions are known for the three regions, it is straight forward [33] to calculate the transmission coefficient for the $m^{th}$ Floquet side band by applying the continuity of $\Psi(x,y,t)$ and $\hat{v}_x\Psi(x,y,t)$ at the two interfaces; $\hat{v}_x$ being the velocity operator.

Finally, the transmission coefficients (given by the ratio of the transmitted current density to the incident one) for the $m^{th}$ side band $(T_m)$ and the total transmission $(T)$ are respectively given by

$$T_m = \frac{\{\alpha k_1^m + \gamma\xi_1^m - \lambda k_1^m(\xi_1^m)^2\}}{\{\alpha k_i^0 + \gamma\xi_i^0 - \lambda k_i^0(\xi_i^0)^2\}}|t_m|^2 \quad \text{and} \quad T = \sum_{m=-\infty}^{m=\infty} T_m \quad (11)$$

where $t_m$ being the ratio of the amplitude of the transmitted wave to that of the incident, $k_i^0$ being the incident wave vector and $\xi_i^0 = \xi_1^m (for\; A_m = 0\; and\; m = 0)$.

Finally, the integration over $k_y$ of the total transmission (T, in eqn.(11)) reproduces the zero temperature conductance (G) [34, 35] under laser assisted condition and is given by

$$G = G_0 \int_{-k_{max}}^{k_{max}} T \, dk_y \quad (12)$$

$k_{max}$ being the maximum value of the y – component of momentum and $G_0$ is a constant having the dimension of conductance.

Results and discussion:

The electronic transport through a rectangular vector potential barrier (RVB) in phosphorene can be thought as equivalent to a pair of spatially displaced delta function magnetic fields directed opposite to each other and perpendicular to the plane of propagation. Such tunneling transport studies have not been reported so far even under static condition. First of all we discuss a few results (vide Figs. 2(a) – 2(c)) under the field (laser) free (FF) condition of the vector barrier. The parameters in all the figures are in dimensionless unit, e.g., energy (E) in $E_0$, length in $L_0 = \sqrt{\frac{\hbar^2}{2m_0 E_0}}$, magnetic vector potential in $A_{m0} = B_0 L_0$, laser frequency in $\omega_0 = \frac{E_0}{\hbar}$ and amplitude of the laser field in $A_{l0} = \frac{\hbar}{eL_0} = \frac{F_0}{\omega_0}$ (where amplitude of the laser electric field is expressed in $F_0$). Typically, for a particilar value $E_0 = 0.1$ eV, the other parameters will be the following; $L_0 = 6.17275$ A$^0$, $A_{m0} = 1.06634$ T-m, $\omega_0 = 15.1924 \times 10^{13}$ Hz, $F_0 = 1.62002 \times 10^{-2}$ V/m and the corresponding laser intensity in $I_0 = 34.8555 \times 10^{12}$ watt/m$^2$. It should be mentioned here that in the present study we have considered the transmission along the x-direction, although the transmission in phosphorene (unlike graphene) is asymmetric in nature along the x and y directions. Figs. 2(a) -2(c) reveals some important features as follows. The transmission is allowed only within certain range of $k_y$ given by $-(k_y)_{max} \leq k_y \leq (k_y)_{max}$. Unlike an electrostatic barrier [17, 36], the laser free transmission coefficient (LFTC) through a RVB is asymmetric with respect to the sign of $k_y$. It is also noted from the Figs. 2(a) and 2(b) that within the transmission window, the RVB is comparatively more transparent for negative values of $k_y$ than the positive one. This feature may be attributed to the fact that the effective barrier height depends on the magnitude $|k_y + A_m|$. The magnitude of the anisotropic transmission decreases systematically with the increase of the magnetic field strength, although the allowed range over $k_y$ remains independent. The asymmetry in the LFTC increases abruptly with the increase in the width (D) of the vector barrier showing a sharp asymmetric peak around the negative end of the $k_y$ axis (vide Figs. 2(a) and 2(b)). Interestingly, for $D = 4$ (Fig. 2(b)) the variation of the LFTC is almost linear with respect to $k_y$ and the normally incident electron suffers maximum changes in transmission with the increase in D. In comparison to a linearly varying vector potential barrier [20], the LFTC through a RVB manifests two interesting differences as follows: the LFTC remains constant within the transmission window in the former case [20], in sharp contrast to a systematic decrease in LFTC with increasing $k_y$ (particularly in the +ve region) for the latter. Further in the former case, the allowed range of transmission depends on the magnitude of the magnetic field unlike the latter (RVB), where the width of the transmission window depends on the incident energy. Regarding the angular dependence of the transmission spectra with respect to the incident energy, it may be noted from the Fig.2(c) that the angular range and the magnitude of the transmission increases with the increase in incident energy. We finally conclude from the above figures that the transparency of the scattering region decreases with the increase in the height and width of the RVB. These results are well justified from energy the conservation relation and are also quantum mechanically consistent.

In order to study the effect of the external laser field on the electron transmission through a phosphorene rectangular vector potential barrier, we have displayed in Figs. 3(a) and 3(b) the transmission coefficients against $k_y$ in case of larger barrier width (D = 5) for different values of the laser parameters, e.g., the intensity (Fig. 3(a)) and the frequency (Fig. 3(b)). It may also be revealed from the above figures that the LFTC (for large D) exhibits an asymmetric resonant peak at the negative end of the $k_y$ axis (for $k_y \sim -1.0$) while with increasing $k_y$, the transmission also vanishes exponentially for positive values of $k_y$. However, in presence of the laser field, the transmission profile changes to an oscillatory nature such that the $T_c$ decreases within the narrow resonant window (vide Fig. 3(a)) but increases elsewhere. The effect of the laser becomes more pronounced with the increase in the laser intensity. The oscillatory behavior is quite justified due to the appearance of the Bessel function in the wave function for the laser irradiated central region. In fact, the resonant like transmission disappears slowly with increasing laser intensity producing the oscillatory pattern with respect to the variation of $k_y$. The allowed angular range of transmission remains almost independent of the intensity of the laser field.

In order to study the effect of the laser frequency on the magneto-transmission for a wider barrier condition, we have depicted in Fig. 3(b) the $T_c$ for ω = 3, 5, 10 and12. For higher frequency (ω = 12), the effect of the laser is only appreciable for positive values of $k_y$, while the resonant feature remains unaffected. In contrast, the angular transmission profile is much more sensitive both qualitatively and quantitatively w.r.t the lower laser frequency, producing an increasingly oscillatory transmission. Thus the overall effect of laser is almost negligible for high frequency and low intensity (vide Figs. 3(a) and 3(b)). This feature may be attributed to the fact that for lower intensity, the incident photon density decreases, while for higher frequency, the probability of photon exchange process decreases.

Interesting features are noted in the transmission spectra when the width of the vector barrier is decreased. For such case, the angular transmission profile shows a systematic suppression with the increase in laser frequency. Thus the study of the transmission spectra for lower width of the vector barrier needs special attention since the suppression of transmission by external fields is highly demanding for proper exploitation of phosphorene in digital applications.

In view of the above discussion, we display in Figs. 4(a) and 4(b) the energy dependence of the transmission spectra for electron through a laser driven RVB of width D = 1 for different values of the laser amplitude and frequency respectively. Since for lower values of the barrier width D, the transmission is appreciable around the normal incidence, we display in Figs. 4(a) and 4(b), the $T_c$ for normal incidence only. In the absence of the laser, the energy variation of $T_c$ exhibits an exponential rise with the increase in incident energy. In the case of irradiation, the laser induced transmission coefficient (LTC) exhibits an overall suppression in addition to the characteristic Fano resonances (FR) at comparatively higher values of the laser intensity (vide Fig. 4(a)). For $A_m = 2$, the considered energy range displays two small asymmetric FR (a sharp maximum followed by a sharp minimum) that arise due to the quantum interference between the discrete and the continuum. It is now well understood that the presence of a tunneling quasi bound state within the structure is responsible for the FR and appears when the coupling between the electron continuum outside the barrier (or well) and the quasi bound state inside the barrier (or well) takes place due to the photon exchange from the external laser field. With the increase in laser intensity, a sharp anti-resonance appears at the expense of two nearby FRs. The origin may probably be attributed to the Feshbach anti-resonance arising due to the coupling between two nearby FRs mediated by photon. The appearance of the sharp anti-resonance at E ~ 0.5 justifies the presence of a zero energy quasi-bound state supported by the field free structure. The occurrence of the sharp anti-resonance and its control by external field is supposed to be quite interesting for the application in switching and sensing nano-devices.

Fig. 4(b) depicts the effect of laser frequency on the photon induced FR spectrum in tunneling through phosphorene based magnetic barrier structure. With a view to giving a special emphasis on the anti-resonant transmission (useful in regard of device fabrication), we consider the laser potential $(A_m) > 1.5$, for which a dip anti-resonance is noted within the energy range considered here. Fig. 4(b) again concludes that the external laser field overall suppresses the magneto-optic transmission through phosphorene nanostructures. Fig. 4(b) further reveals that with the increase in laser frequency, the position of anti-resonance moves towards the higher energy and in accordance with the energy conservation principle. Further, the magnitude of $T_c$ at the dip increases with the increase in frequency indicating that lower frequency ($\omega \sim 0.5$) will be suitable to produce maximum suppression in tunneling current for better switching action.

Finally, in order to study the effect of the external laser on the tunneling conductance for a phosphorene vector barrier, we display the variation of dimensionless conductance ($G/G_0$) on the barrier width at different values of the laser amplitude and frequency in Figs. 5(a) and 5(b) respectively. Under the laser free condition, the conductance decreases monotonically with increasing barrier width. While with irradiation, the conductance of the system depends strongly on the frequency and intensity of the laser. The laser assisted conductance is suppressed appreciably with a monotonic fall when the barrier ($D$) width becomes smaller. On the other hand, for larger width, the qualitative nature of the conductance changes remarkably. Instead of a monotonic decrease, the conductance in this case becomes highly oscillatory with increasing barrier width and the laser intensity (vide Fig. 5(a)), while the magnitude of the conductance becomes higher than the field free one. Further, the minimum of the conductance profile decreases with the decrease in laser frequency (Fig. 5(b)).

Conclusion:

The present study concerns the electronic transport through a pair of delta function magnetic barrier oriented in such a way that the corresponding vector potential is uniform along the zig-zag direction but steplike along the armchair direction. The magnitude of transmission is found to suppress with increasing the height of the vector barrier. Transmission profile displays the collimated behavior when the width of the barrier is comparatively high or the energy of the charge carrier is low. The effect of the laser is to quench the collimated transmission particularly because of the presence of the Bessel function that makes the transport oscillatory in nature. Transport profile exhibits characteristic Fano resonances for low intensity laser. However, with increasing intensity, the FR turns into a sharp anti-resonance due to the destructive quantum interference effect. Finally, the magneto-optical conductance decreases remarkably with the increase (decrease) of the intensity (frequency) of the laser. Thus the presence of the magnetic vector potential promotes the light-matter interaction even when the polarization of the linearly polarized laser is oriented along the zig-zag direction. The present study confirms that the presence of the magnetic vector potential on phosphorene nanostructure makes it more flexible towards the photo-absorption and photo-detection.

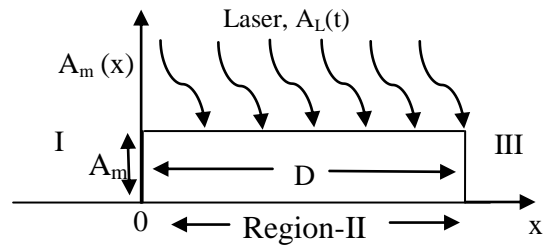

Fig.1: Schematic diagram for a rectangular magnetic vector potential profile $A_m(x)$ corresponding to a pair of oppositely directed delta function magnetic barriers separated by a distance '*D*'. Laser is applied in region II.

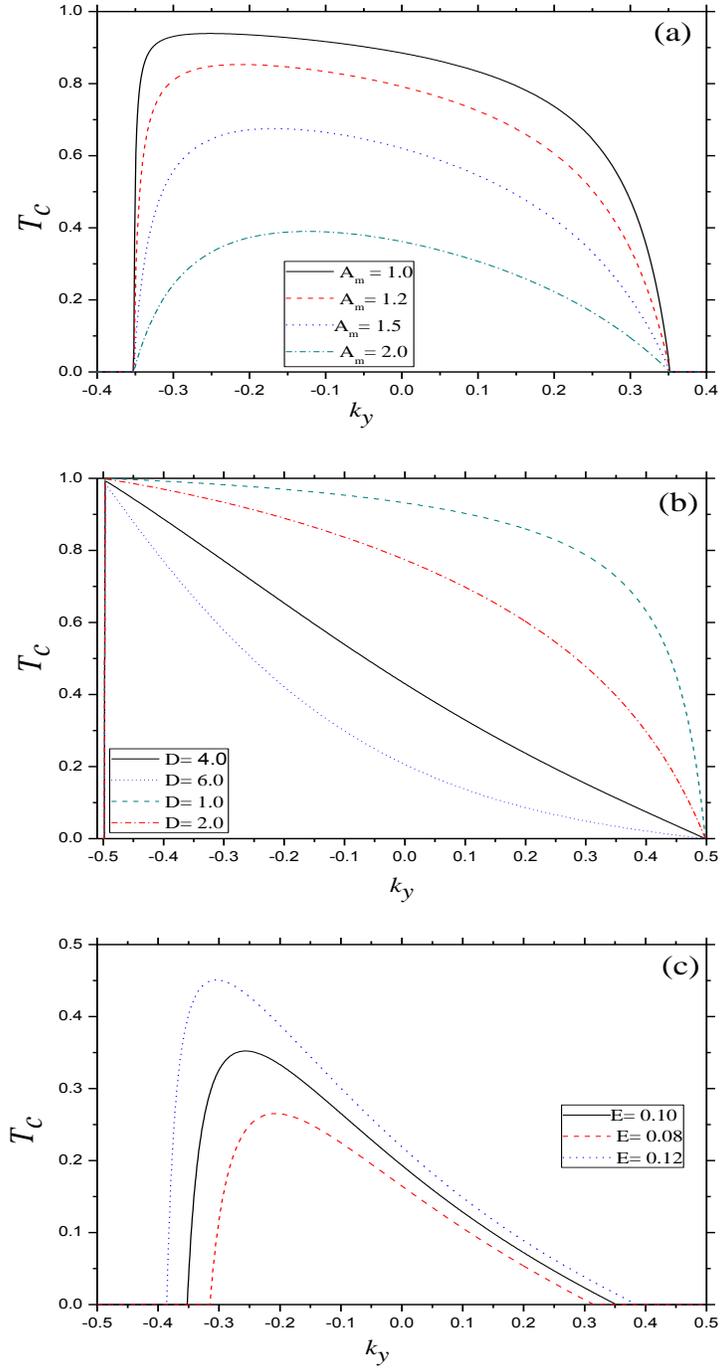

Fig.2: Laser free transmission coefficient ($T_c$) as a function of wave vector $k_y$. (a) For 'D' = 1, 'E' = 0.1 and $A_m$= 1.0, solid (black) line; = 1.2, dash (red) line; =1.5, dot (blue) line; = 2.0, dash-dot (green) line. (b) For $A_m$= 1.0, 'E' = 0.2 and 'D' = 6.0, dot (blue) line; = 4, solid (black) line; = 2, dash-dot (red) line; = 1.0 dash (green) line. (c) For 'D' = 5.0, $A_m$= 1.0 and 'E'= 0.08, dot (red) line; = 0.1, solid (black) line; = 0.12, dash (red) line.

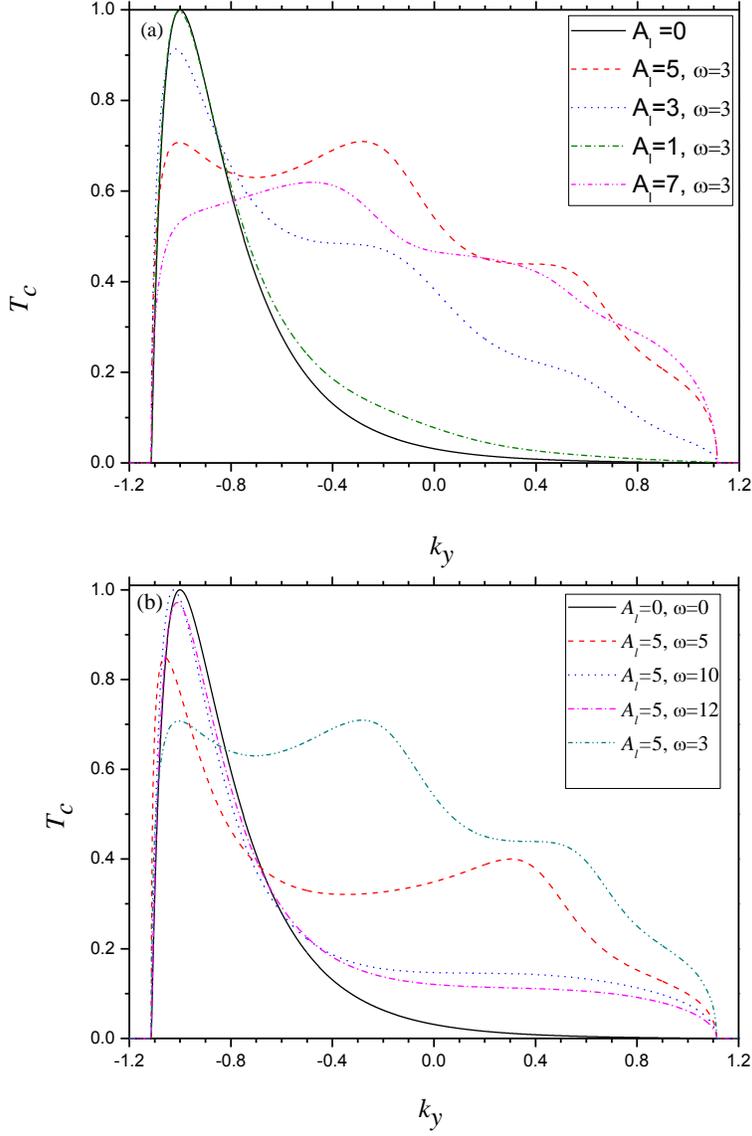

Fig.3: Laser assisted transmission coefficient ($T_c = \sum_m T_m$) as a function of wave vector $k_y$. For 'D' = 5, 'E' = 1.0 and $A_m$ = 2.0. Solid (black) line corresponds to laser free case. (a) For 'ω' = 3.0 and $A_l$ = 1, dash-dot (green) line; = 3, dot (blue) line; = 5, dash (red) line; = 7, dash-dot-dot (purple) line. (b) For $A_l$ = 5.0 and 'ω' = 3.0, dash-dot-dot (green) line; = 5.0, dash (red) line; = 10.0, dot (blue) line; = 12.0, dash-dot (purple) line.

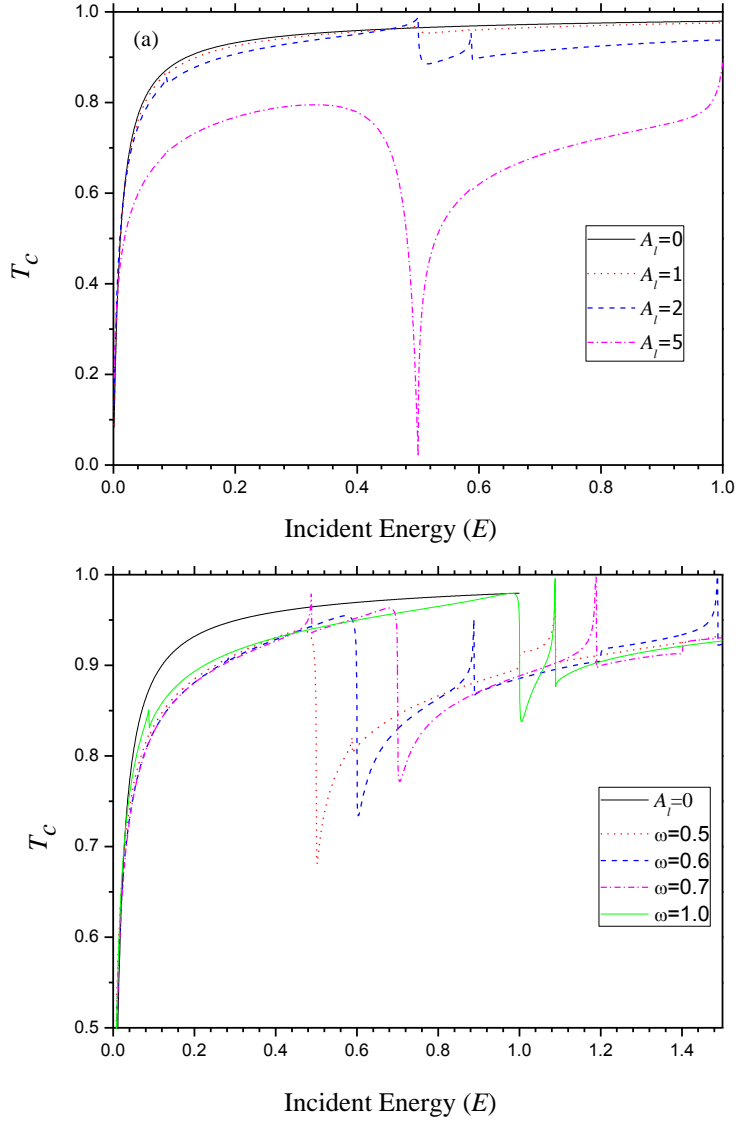

Fig.4: Same as Fig.3 but as a function of incident energy. For 'D' = 1.0, $A_m$ = 1.0 and $k_y$ = 0. Solid (black) line corresponds to laser free case. (a) For 'ω' = 0.5 and $A_l$ = 1, dot (red) line; = 2, dash (blue) line; = 5, dash-dot (purple) line. (b) For $A_l$ = 3 and 'ω' = 0.5, dot (red) line; = 0.6, dash (blue) line; = 0.7, dash-dot (purple) line; = 1.0, dash-dot-dot (green) line.

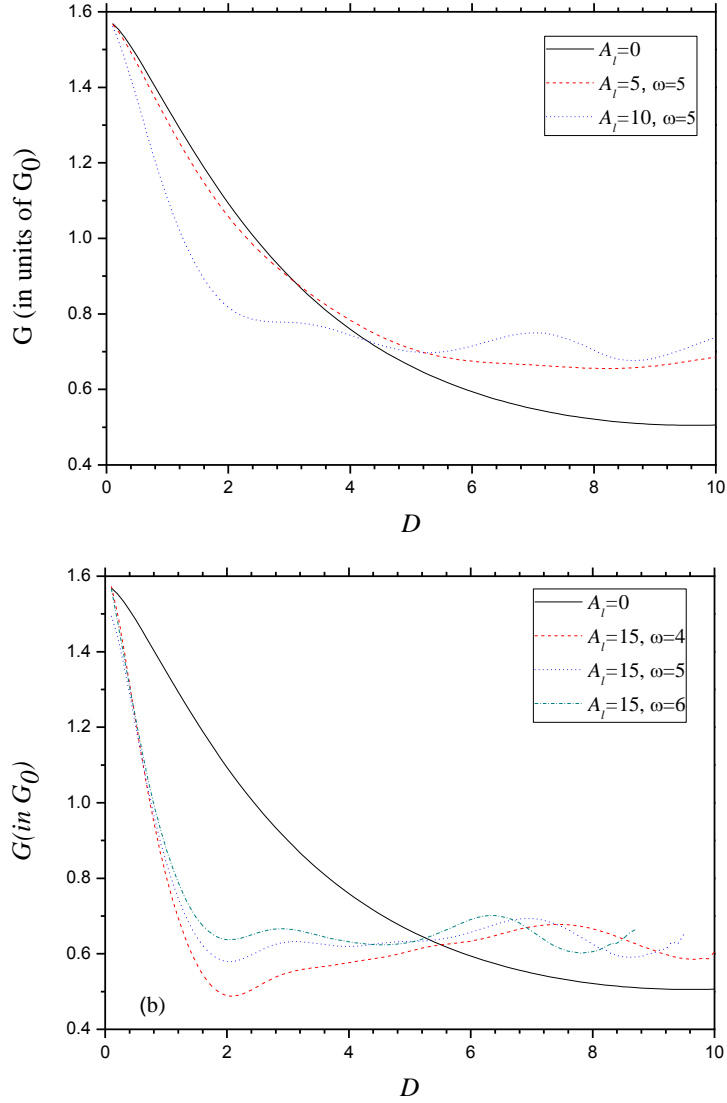

Fig.5: Conductance 'G' (in units of $G_0$) plotted as function of barrier width for 'E' = 0.5 and $A_m$ = 1. Solid (black) line corresponds to laser free case. (a) For 'ω' = 5 and $A_l$ = 5, dash (red) line; = 10, dot (blue) line. (b) For $A_l$ = 15 and 'ω' = 4, dash (red) line; = 5, dot (blue) line; = 6, dash-dot (green) line.